\makeatletter
\declare@file@substitution{revtex4-1.cls}{revtex4-2.cls}
\makeatother
\documentclass[nolinenumbers]{aastex63}

\shorttitle{Evidence for Abiotic Dimethyl Sulfide}
\shortauthors{Hänni et al.}
\begin{document}

\title{Evidence for Abiotic Dimethyl Sulfide in Cometary Matter}

%\author[0000-0002-0786-7307]{Greg J. Schwarz}
%\affiliation{American Astronomical Society \\
%1667 K Street NW, Suite 800 \\
%\collaboration{1}{(AAS Journals Data Scientists collaboration)}

\correspondingauthor{Nora Hänni}
\email{nora.haenni@unibe.ch}

\author{Nora Hänni}
\affiliation{Physics Institute, Space Research \& Planetary Sciences, University of Bern, Sidlerstrasse 5, CH-3012 Bern, Switzerland.}

\author{Kathrin Altwegg}
\affiliation{Physics Institute, Space Research \& Planetary Sciences, University of Bern, Sidlerstrasse 5, CH-3012 Bern, Switzerland.}

\author{Michael Combi}
\affiliation{Department of Climate and Space Sciences and Engineering, University of Michigan, Ann Arbor, MI, USA.}

\author{Stephen A. Fuselier}
\affiliation{Space Science Division, Southwest Research Institute, San Antonio, TX, USA.}
\affiliation{Department of Physics and Astronomy, The University of Texas at San Antonio, San Antonio, TX, USA.}

\author{Johan De Keyser}
\affiliation{Royal Belgian Institute for Space Aeronomy, BIRA-IASB, Brussels, Belgium.}

\author{Niels F. W. Ligterink}
\affiliation{Physics Institute, Space Research \& Planetary Sciences, University of Bern, Sidlerstrasse 5, CH-3012 Bern, Switzerland.}
\affiliation{Faculty of Aerospace Engineering, Delft University of Technology, Delft, The Netherlands.}

%\author{Daniel R. Müller}
%\affiliation{Physics Institute, Space Research \& Planetary Sciences, University of Bern, Sidlerstrasse 5, CH-3012 Bern, Switzerland.}

\author{Martin Rubin}
\affiliation{Physics Institute, Space Research \& Planetary Sciences, University of Bern, Sidlerstrasse 5, CH-3012 Bern, Switzerland.}

\author{Susanne F. Wampfler}
\affiliation{Center for Space and Habitability, University of Bern, Gesellschaftsstrasse 6, CH-3012 Bern, Switzerland.}

\begin{abstract} %%%250 words limit
Technological progress related to astronomical observatories such as the recently launched James Webb Space Telescope (JWST) allows searching for signs of life beyond our Solar System, namely in the form of unambiguous biosignature gases in exoplanetary atmospheres. The tentative assignment of a 1-2.4$\sigma$ spectral feature observed with JWST in the atmosphere of exoplanet K2-18b to the biosignature gas dimethyl sulfide (DMS; sum formula C$_2$H$_6$S) raised hopes that, although controversial, a second genesis had been found. Terrestrial atmospheric DMS is exclusively stemming from marine biological activity and no natural abiotic source has been identified -- neither on Earth nor in space. Therefore, DMS is considered a robust biosignature. Since comets possess a pristine inventory of complex organic molecules of abiotic origin, we have searched high-resolution mass spectra collected at comet 67P/Churyumov-Gerasimenko, target of the European Space Agency's \textit{Rosetta} mission, for the signatures of DMS. Previous work reported the presence of a C$_2$H$_6$S signal when the comet was near its equinox but distinction of DMS from its structural isomer ethanethiol remained elusive. Here we reassess these and evaluate additional data. Based on differences in the electron ionization induced fragmentation pattern of the two isomers, we show that DMS is significantly better compatible with the observations. Deviations between expected and observed signal intensities for DMS are $<1\sigma$, while for ethanethiol they are ~2-4$\sigma$. The local abundance of DMS relative to methanol deduced from these data is $(0.13\pm0.04)\%$. Our results provide the first evidence for the existence of an abiotic synthetic pathway to DMS in pristine cometary matter and hence motivate more detailed studies of the sulfur chemistry in such matter and its analogs. Future studies need to investigate whether or not the present inference of cometary DMS could provide an abiotic source of DMS in a planetary atmosphere.
\end{abstract}

\keywords{comets:general -- comets: individual: 67P/Churyumov-Gerasimenko -- instrumentation: detectors -- methods: data analysis}

%%%%%%%%%%%%%%%%%%%%%%%%%%%%%%%%%%%%%%%%%%%%%%%%%%%%%%%%%%%%%%%%%%%%%%%

\section{Introduction} \label{sec:intro}
Based on data collected during an Earth fly-by by the National Aeronautics and Space Administration (NASA) spacecraft \textit{Galileo}, \citet{sagan1993} proposed that unique molecular indicators of carbon-based life on Earth could be used to search for and detect extraterrestrial life. Today, those indicators are often referred to as biosignatures or sometimes as biomarkers, the latter term being borrowed from the field of medicine. The use of biosignature gases is especially interesting for the search for life in environments like exoplanets that are not accessible by space missions for \textit{in-situ} studies. Currently, potential life on exoplanets is only detectable with remote sensing spectroscopic techniques using ground- or space-based observatories. A recent example is NASA's James Webb Space Telescope, capable of investigating exoplanetary atmospheres in the near-to-mid-infrared spectral range in unprecedented detail. Taking advantage of the constantly increasing resolving power, wavelength coverage, and sensitivity of exoplanet observatories, \citet{seager2016} compiled a list of terrestrial atmospheric molecules that are produced by biological activity and hence could be employed as biosignatures. Crucial properties of such biosignature gases are, besides volatility, abundance, and stability under exoplanet-like conditions, unique spectral features and the absence of false positives. Beyond the candidates already suggested by \citet{sagan1993}, molecular oxygen (O$_2$; produced by photosynthesis) and methane (CH$_4$; produced by biology, for example through methanogenesis), the list in \citet{seager2016} also mentions the simple sulfur-bearing molecule dimethyl sulfide (DMS; C$_2$H$_6$S). Methane can be produced by both biotic and abiotic chemical processes, including methanogenic bacteria \citep{taubner2018} and oceanic hydrothermal vents \citep{horita1999}. This may be relevant, for instance, for Saturn's icy moon Enceladus \citep{bouquet2015,waite2017}. So far, for DMS, no false positives are known. However, \citet{raulin1975} have shown experimentally that electrical discharge in a gaseous mixture of hydrogen sulfide (H$_2$S) and methane can lead to the abiotic formation of DMS.
DMS falls into the class of low-abundance but unique biosignatures that originate from secondary reactions, i.e., reactions that are not related to energy production or biomass build-up \citep{seager2016}. On Earth, the most relevant and best-studied source of DMS is the degradation of dimethylsulfoniopropionate (DMSP; C$_5$H$_{10}$O$_2$S), a sulfur-bearing compound produced by eukaryotic marine phytoplankton and a few higher plants \citep{andreae1983,pilcher2003}. DMS, among a handful of small molecules, has been discussed as potentially observable indicator of aquatic life on habitable exoplanets with a hydrogen-dominated atmosphere \citep{domagal-goldman2011,seager2013,catling2018,schwieterman2018} and, more recently, on a specific class of ocean-covered exoplanets with a hydrogen-rich atmosphere, referred to as Hycean worlds \citep{madhusudhan2021,rigby2024}. Shortly after, \citet{madhusudhan2023} reported the tentative detection of DMS in the atmosphere of exoplanet K2-18b, a candidate Hycean world, with the James Webb Space Telescope using transit spectroscopy. Despite the DMS detection's confidence was only 2.4$\sigma$/$\approx$1$\sigma$/none (depending on how the relative detector offset was accounted for), this work received world-wide attention as it featured DMS as a potential indicator of marine biological activity on this planet, adding though that more conclusive observational constraints on the DMS atmospheric abundance are needed. Better constrained abundances of atmospheric species could also underpin the classification of K2-18b as a Hycean world, which is being debated \citep{wogan2024,shorttle2024,glein2024}. Generally, the search for extraterrestrial life should rely on multiple lines of evidence, cf., e.g., \citet{catling2018} or \citet{schwieterman2018}.\\
\indent We take a different path in this debate by asking if DMS is present in abiotic cometary matter. Comets are known to be rich in refractory and volatile organic matter \citep{sandford2008,bardyn2017,altwegg2019} that is thought to be well-preserved since the earliest times of our Solar System \citep{bockelee-morvan2000,drozdovskaya2019,lopez-gallifa2024}, and hence, are a relevant point of reference when it comes to abiotic chemical complexity and diversity.\\
\indent For this work, we revisited the organosulfur inventory of comet 67P \citep{calmonte2016,mahjoub2023} and searched for evidence of the presence of DMS, and hence, for an abiotic source of this molecule. Comet 67P was the target of the European Space Agency's (ESA's) \textit{Rosetta} mission. It was the first comet to be studied from up-close for an extended time period of two years as it passed through perihelion. The \textit{Rosetta} Orbiter Instrument for Ion and Neutral Analysis \citep[ROSINA; ][]{balsiger2007} with its high-resolution Double Focusing Mass Spectrometer (DFMS) was collecting unique data that unveiled this comet's chemical inventory in unprecedented detail \citep{rubin2019b,altwegg2019}. Many surprising chemical species were detected, from molecular oxygen \citep{bieler2015} to ammonium salts \citep{haenni2019,altwegg2020,poch2020,altwegg2022}, and including a plethora of complex organic molecules \citep{schuhmann2019a,schuhmann2019b,haenni2022,haenni2023}. \citet{calmonte2016} mentioned the presence of a signal at the exact mass of C$_2$H$_6$S in multiple ROSINA/DFMS mass spectra. However, due to overlap with the nearby C$_5$H$_2$ signal as well as structural isomerism of DMS (CH$_3$--S--CH$_3$) and ethanethiol (CH$_3$--CH$_2$--SH), these authors did not make conclusive statements regarding the presence or absence of DMS in comet 67P. Here, we revisit data from \citet{calmonte2016} and, thanks to a more precisely constrained mass scale, we arrive at a spectral deconvolution that allows the distinction of DMS and ethanethiol. Technical and methodological details are described in Section \ref{sec:data}, while Section \ref{sec:res-disc} presents and discusses our results. Section \ref{sec:concl} summarizes our findings, stating how they motivate further research on the sulfur chemistry in the early Solar System, and discusses potential implications for exoplanetary science.\\

%(62.0185 Da for the molecular ion)
%(62.0151 Da for the molecular ion)

\section{Instrument and Method}\label{sec:data}
ROSINA/DFMS data collection and reduction have been described in detail by \citet{leroy2015} and \citet{calmonte2016}. In the following, we will therefore limit ourselves to the presentation of the most relevant aspects.

\subsection{Data Acquisition with DFMS}\label{subsec:dfms}
DFMS, together with a time-of-flight mass spectrometer and a pressure gauge making up the ROSINA instrument suite \citep{balsiger2007}, was designed to study the volatile species in comet 67P's coma. Cometary neutrals entering the DFMS' ionization chamber were subjected to electron ionization (EI) using a 45 eV electron beam. Ions were subsequently extracted from the ion source and passed through an electrostatic analyzer and a permanent magnet, where the electric and magnetic fields led to separation of the species according to their mass-per-charge ratio (\textit{m/z}). The ions were detected by a stack of two micro channel plates mounted in a Chevron configuration. The released charge was collected on two rows of position sensitive linear electron detector array anodes. Scanning through the DFMS' mass range was done by stepping from integer mass-per-charge to integer mass-per-charge sequentially and took approximately 30 s per mass spectrum, of which 20 s was the integration time. We note that the micro channel plate detector works in analog mode. This means that the charge deposited on the anodes during integration, i.e., the detector signal, depends not only on the number of impacting ions, but also on their energy, their structure and size, as well as on the detector gain. The detector gain is determined by the set gain step (i.e., the voltage used to amplify the detector signal) and the individual pixel gain. The latter factor relates to the inhomogeneous detector aging that affects the detector on timescales on the order of weeks and months. The pixel gain is corrected by frequently measured correction factors. Detailed information on pixel gain corrections have been presented in \citet{leroy2015} and \citet{dekeyser2019} or in \citet{calmonte2015} and \citet{schroeder2020}. In subsection \ref{subsec:red}, we estimate the errors, including uncertainties introduced by the detector gain.\\
\indent The main data set analyzed for this work was collected on 15 March 2016 between 03:30 and 04:14 UTC, during comet 67P's outbound equinox, and covers the mass range \textit{m/z}=13--100. That day, the most intense C$_2$H$_6$S signal was observed, based on the DFMS data evaluated so far. DFMS achieved a mass resolution of \textit{m/$\Delta$m}=3000 at 1\% of the peak height at \textit{m/z}=28 \citep{balsiger2007}, which is exceptional for space instruments and allows searching for the fingerprints of specific molecules in a complex analyte mixture as we explain in detail the in following two subsections.

\subsection{Electron Ionization-Induced Fragmentation}\label{subsec:fragment}
\begin{figure}[ht]
   \centering
   \includegraphics[width=8cm]{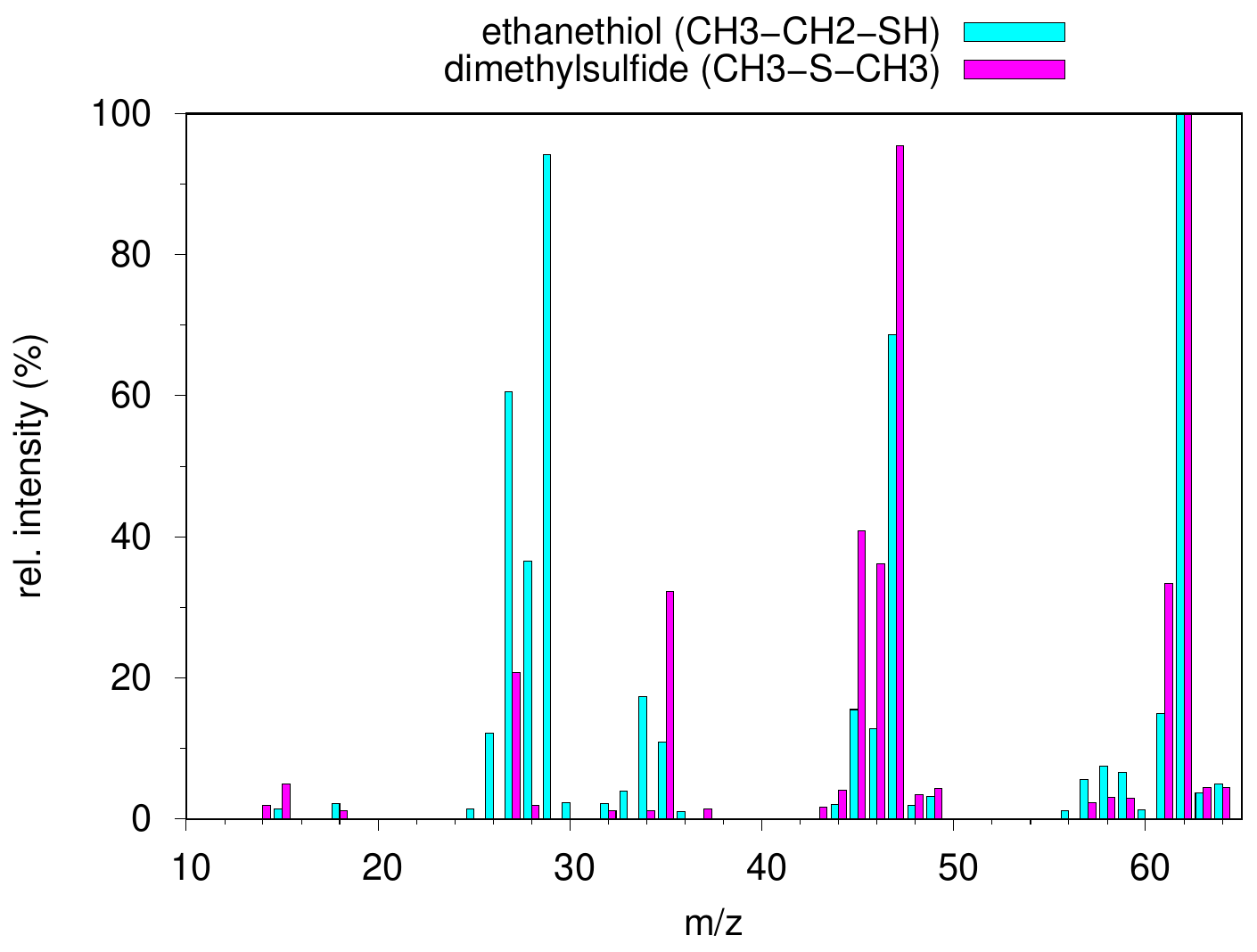}
      \caption{Juxtaposed NIST reference fragmentation patterns of the two structural isomers DMS and ethanethiol \citep{NIST}. The relative intensity (rel. intensity; normalized to the highest signal in the spectrum) is plotted as a function of mass-per-charge (\textit{m/z}).}
         \label{fig:frag}
 \end{figure}

EI mostly yields singly charged cations of the analyte molecules, which we refer to as molecular ions (M), or fragments thereof. Fragmentation accompanying the EI process produces the so-called fragmentation pattern of an analyte species, which is the characteristic fingerprint of signals of all charged species, parents and fragments. If the gaseous analyte is a complex mixture, such as that from a cometary coma, the fragmentation patterns of all species that are simultaneously ionized in the instruments' ionization chamber overlap and the sum of their fragmentation patterns is detected. As it is not possible to calibrate all volatile organic molecules on the DFMS laboratory twin instrument and the DMFS calibration facility is not equipped for handling corrosive or toxic species, like for instance DMS, we used reference fragmentation patterns from the National Institute of Standards and Technology (NIST) chemistry web-book EI-MS data base \citep{NIST} for this work. Fragmentation patterns measured with DFMS are known to deviate little from those in the NIST data base, see, e.g., \citet{schuhmann2019a,schuhmann2019b}. This deviation is mostly due to the different ionization energies -- DFMS uses 45 eV electrons while NIST standard spectra are usually measured using 70 eV electrons. The ionization cross-section of DMS changes little between 45 and 70 eV \citep{kaur2015}. By trend, lower ionization energies may slightly enhance the relative abundance of the parent with respect to the fragments. Also, differences in instrument geometries may play a role. Here, we do not account for those deviations. However, calibration experiments performed by \citet{schuhmann2019a,schuhmann2019b} and in the framework of several PhD theses \citep{calmonte2015,schroeder2020,schuhmann2020} archived together with the ROSINA data on ESA's Planetary Science Archive (PSA) and NASA's Planetary Data System (PDS) show that they are generally small, normally below 10\% relative intensity. Figure \ref{fig:frag} juxtaposes the NIST fragmentation patterns of DMS and its structural isomer ethanethiol, the topic of this work, normalized to the highest signal. The most intense signal in both fragmentation patterns is the M signal on \textit{m/z}=62. Both isomers also produce a relevant M-H signal on \textit{m/z}=61, which for DMS is more pronounced. The lower-mass fragments, however, are distinct: DMS favorably loses a methyl group, leading to a strong group of signals on \textit{m/z}=45-47, the M-CH$_3$ fragment being the one on \textit{m/z}=47. Ethanethiol may also lose a methyl group, but more favorably loses the thiol function to produce a set of strong signals on \textit{m/z}=27-29. Below, we discuss which features are used to distinguish the two molecules based on ROSINA/DFMS mass spectra.\\

\subsection{Data Reduction}\label{subsec:red}
\begin{figure}[ht]
  \centering
  \includegraphics[width=8cm]{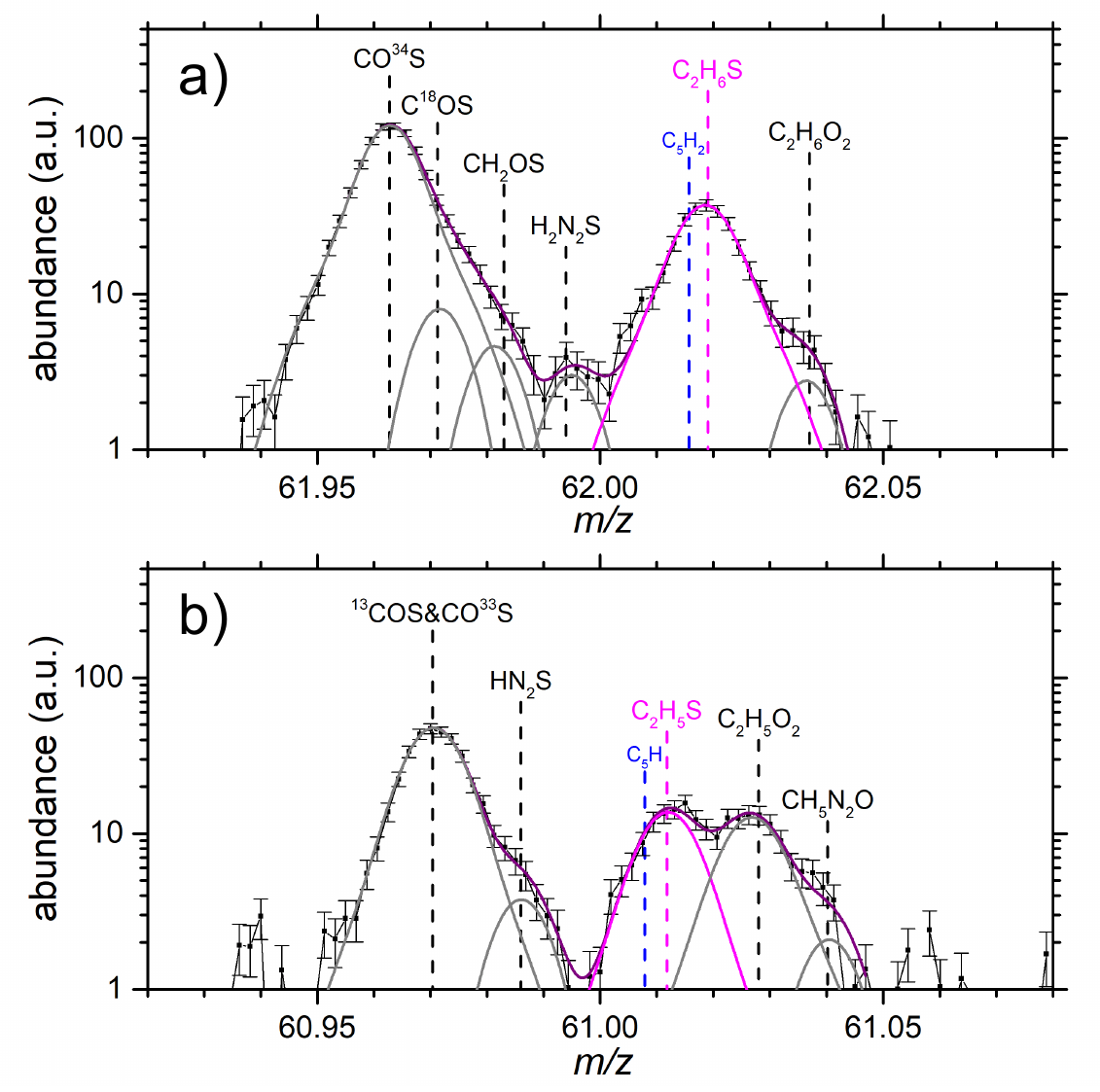}
     \caption{DFMS mass spectra collected on 15 March 2016 around \textit{m/z} equal to 62 (a) and 61 (b). The measured intensities per pixel are plotted with black squares, in arbitrary units (a.u.), and with their statistical uncertainties. The peaks are fitted with Gaussian functions. We show the individual Gaussians in gray, highlighting those representing the relevant S-bearing species in magenta, and the overall fit in purple. The exact mass positions of the different chemical species are indicated by dashed lines. The standard deviation of the fitted peak position is $\approx 6\times 10^{-4}$.}
        \label{fig:spec}
\end{figure}

Figure \ref{fig:spec} shows the mass spectra registered around the nominal masses \textit{m/z}=62, where we expect the molecular ion of DMS (C$_2$H$_6$S=M), and \textit{m/z}=61, where, according to NIST \citep{NIST}, we expect an abundant DMS fragment (C$_2$H$_5$S=M-H). Here, we describe the data reduction routine to extract the relative intensities from ROSINA/DFMS mass spectra:\\
\indent In a first step, the absolute peak position and, consequently, the mass scale in the mass spectrum, has to be defined via one known species-peak pair. Frequently observed continuation of homologous series in adjacent mass spectra (e.g., C$_2$H$_5$S and C$_2$H$_6$S in Figure \ref{fig:spec}) as well as regularly measured and abundant cometary species like H$_2$O, CO, CO$_2$, COS, or CS$_2$ are helpful to verify the mass scale. The mass scale around \textit{m/z}=61 and 62 is very well constrained by the strong signals of COS and its isotopologue OC$^{34}$S on \textit{m/z}=60 and 62, respectively, as well as by the SO$_2$ and S$_2$ signal on \textit{m/z}=64. The pixel zero (p$_0$), i.e., the pixel of the commanded integer mass, shifts only little ($\approx$0.3 pixel) from one mass scan to the next.

\indent Subsequently, the signal strengths must be extracted via peak fitting, using a Gaussian peak profile on a linear background, see, e.g., \citet{leroy2015}. A good fit is usually achieved by combining an intense and narrow first Gaussian with a less intense and broader second Gaussian with fixed height and width ratios \citep{dekeyser2019}. Here, we used a double-Gaussian peak profile for signals with intensities above 10 a.u. and a single-Gaussian one for smaller signals where the second Gaussian disappears in the background. It can be seen from Figure \ref{fig:spec} that signals from multiple species are usually present in one mass spectrum. Thanks to the high mass resolution of DFMS, it is often possible to distinguish species that have only small differences in the exact masses. The fitted peak position for the signal attributed to C$_2$H$_6$S is 62.0185 $\pm$ 0.0005 (the exact mass being at \textit{m/z}=62.0190) and for the signal attributed to C$_2$H$_5$S it is 61.0116 $\pm$ 0.0008 (the exact mass being at \textit{m/z}=61.0112). Hence, both peak positions lie within 1$\sigma$ from the expected peak positions of C$_2$H$_6$S and C$_2$H$_5$S. The next closest possible species, C$_5$H$_2$ and C$_5$H, have a mass difference of ~7$\sigma$ and ~4.5$\sigma$, respectively, which renders their contributions negligible. The fit error for the peak area is 3.6\% for C$_2$H$_6$S and 8\% for C$_2$H$_5$S. The reduced $\chi^2$ for the overall fit is 0.998 for \textit{m/z}=62 and 0.992 for \textit{m/z}=61.\\
\indent Sometimes, corrections for a drift in the signal strength during the time of the data collection are necessary as a change in cometary outgassing activity potentially distorts the observed fragmentation patterns. However, no such correction was implemented as the isotopic ratios of CO$_2$ and COS match the expectations \citep{haessig2017,calmonte2017}, which indicates stable coma conditions on the time scales relevant for this investigation. Also, the water signal on \textit{m/z}=18, additionally scanned first and last in every mode to track density fluctuations, shows $<3\%$ variation. In addition, the instrument sensitivity changes as a function of mass and varies for the different species. We accounted for the mass-dependent sensitivity change according to the description, e.g., in \citet{leroy2015}. As the species-dependent sensitivities are not known, it is not possible to derive the signal strength as number of ions per time, and we show it in arbitrary units (a.u.). However, N$_2$ gas was used to calibrate the DFMS detector and the established conversion factor was applied to allow a quantitative interpretation of the resulting numbers as N$_2$$^+$ ion equivalents. The estimated errors in the signal intensities of the C$_n$H$_m$S ($n$ and $m$ being the stoichiometric coefficients of C and H, respectively) species targeted in this work are $\approx$15\%. This estimation for the overall errors includes: (1) a statistical error (negligible for the signals under consideration except for the very low abundant ones like C$_2$H$_3$S), (2) a fit error, and (3) an error related to different signal gains of the individual pixels. Errors related to the detector gain cancel out as all species were measured on the highest gain step. The non-negligible statistical error for low-abundant species increases the respective overall error to be 20\%. 20\% overall error was also estimated for CH$_3$S on \textit{m/z}=47. While the counting statistics is good for CH$_3$S, its exact mass is very close to $^{13}$C$^{16}$O$^{18}$O, an isotopologue of the abundant cometary volatile CO$_2$, and no distinction by fitting is possible. The observed signal was hence corrected by subtracting 1.2\% of the signal of $^{12}$C$^{16}$O$^{18}$O on \textit{m/z}=46 based on the carbon isotope ratio measured in CO$_2$ \citep{haessig2017}. The latter is separable from CH$_2$S by fitting. The C$_2$H$_4$S signal on \textit{m/z}=60 has an even larger overall error of ~50\% as this species is fully in the slope of the large peak of the main COS isotopologue which increases the fit's uncertainties considerably.\\

\section{Results and Discussion}\label{sec:res-disc}
\begin{figure}[ht]
   \centering
   \includegraphics[width=8.5cm]{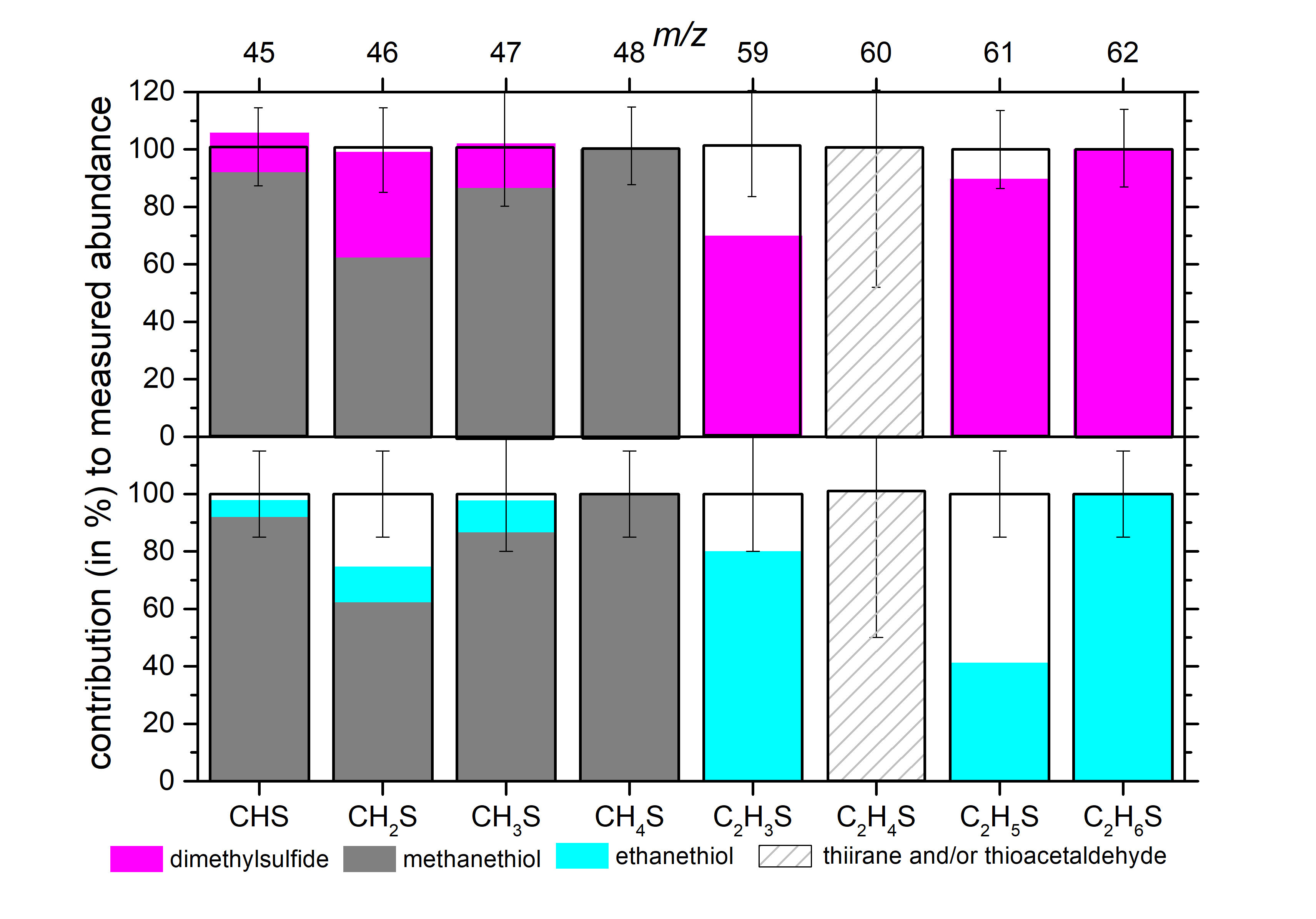}
      \caption{Measured versus expected intensities for the two combinations methanethiol + DMS (top) and methanethiol + ethanethiol (bottom). Color-coded are the individual contributions in percent of the observed intensity. Reference spectra have been taken from NIST \citep{NIST} and were scaled to C$_2$H$_6$S on \textit{m/z}=62 for both DMS and ethanethiol and to CH$_4$S on \textit{m/z}=48 for methanethiol as those species can be unambiguously assigned to the respective molecular ions. The hatched contribution to C$_2$H$_4$S on \textit{m/z}=60 can be assigned to thiirane and/or thioacetaldehyde (parent), fragments of which we do not indicate as no reference spectrum is available for thioacetaldehyde. Details are given in the main text.}
         \label{fig:deconv}
\end{figure}

The fits presented in Figure \ref{fig:spec} and discussed above allow unambiguous association of the observed signals to C$_2$H$_6$S and C$_2$H$_5$S, respectively, and show that the contribution of the pure hydrocarbon species (C$_5$H$_2$ and C$_5$H) are negligible. The corrected observed intensities (in a.u.) are now compared to the reference fragmentation patterns from NIST. Figure \ref{fig:deconv} shows the observed intensities of organosulfur species between \textit{m/z}=45 and 62 and how they compare to the expected intensities \citep{NIST} for the possible combinations of parent molecules. The most striking difference between the two isomers, DMS and ethanethiol, manifests itself with regard to the C$_2$H$_5$S signal at \textit{m/z}=61, corresponding to M-H of both candidates. While fragmentation of DMS explains the observed C$_2$H$_5$S signal within 1$\sigma$ error margins (top panel), the expected ethanethiol fragment intensity is off from the observed signal by more than 3$\sigma$ (bottom panel). The same findings result from the investigation of additional mass spectra collected on the same day (15 March 2016) at different times as well as two days later (17 March 2016). In the appendix, Figure \ref{fig:A1} plots the additional data and Table \ref{tab:A1} lists the derived C$_2$H$_5$S/C$_2$H$_6$S ratios which are compatible, within error limits, with DMS but not with ethanethiol.\\
The observed C$_2$H$_3$S signal on \textit{m/z}=59, corresponding to approximately 8\% of the C$_2$H$_6$S signal on \textit{m/z}=62, is only partially due to the M-3H fragment of DMS or ethanethiol, respectively. The unexplained intensity most likely indicates the presence of an additional low-abundant parent with a molecular sum formula of C$_2$H$_4$S and a nominal mass of 60 Da that produces a C$_2$H$_3$S (M-H) fragment on \textit{m/z}=59. Unfortunately, the C$_2$H$_4$S signal is hidden below the flank of the strong COS signal and can only be retrieved from the data with an uncertainty as large as 50\% (hatched bar in Figure \ref{fig:deconv}). Attributing the C$_2$H$_4$S signal to either thiirane or thioacetaldehyde (or a mixture of both), the two structural isomers of C$_2$H$_4$S, likely explains the C$_2$H$_3$S intensity yet unexplained by DMS or ethanethiol. While the thiirane fragmentation pattern from \citet{NIST} is compatible with the residual intensities in Figure \ref{fig:deconv} (notably DMS residual intensities seem to fit better than those from ethanethiol), the fragmentation pattern of thioacetaldehyde is unknown. Similar to aldehydes, preferred loss of the aldehyde hydrogen atom is likely for thioacetaldehyde, which might lead to strong M and M-H signals too. It has to be stated clearly that, due to the large error margins of the signals and the missing fragmentation information for thioacetaldehyde, no conclusions can be drawn from the C$_2$H$_3$S signal as to which of the two C$_2$H$_6$S isomers produces a better match. Alternatively, heavier organosulfur species might yield C$_2$H$_3$S as a fragment, see further discussion of heavier organosulfur species below.\\
Last but not least, also the observed signals on \textit{m/z}=45-47 favor the presence of DMS over that of ethanethiol. While these signals are mostly attributed to methanethiol (CH$_4$S) and its fragments (gray bars in Figure \ref{fig:deconv}) and contributions of the C$_2$H$_6$S isomers are minor, the CH$_2$S fragment on \textit{m/z}=46 leaves sufficient residual intensity after subtraction of the methanethiol contribution to be distinctive: Again, the DMS contribution matches the missing intensity almost perfectly ($<1\sigma$ difference) while the ethanethiol contribution does not (approximately 2$\sigma$ difference).\\

\indent Based on Figure \ref{fig:deconv} as well as Figure \ref{fig:A1} and Table \ref{tab:A1} in the appendix, DMS consistently reproduces the observed intensities together with methanethiol much better than ethanethiol. However, fragmenting heavier organosulfur species could contribute to the signals between \textit{m/z}=45 and 61, especially to \textit{m/z}=61 (C$_2$H$_5$S) and 46 (CH$_2$S), and explain the intensity that is left unexplained in Figure \ref{fig:deconv} (bottom). We ruled out this possibility based on the following line of reasoning:

\begin{enumerate}
  \item The signal of fully saturated C$_2$H$_6$S on \textit{m/z}=62 can only indicate DMS or ethanethiol or a combination of both, but it cannot be a charge retaining fragment of a heavier organosulfur as such fragments would be unsaturated. Approximately ~40\% of the observed intensity on C$_2$H$_5$S on \textit{m/z}=61 cannot be explained if all the C$_2$H$_6$S signal on \textit{m/z}=62 is associated to ethanethiol. If there are no heavier organosulfur parent species whose fragments could explain this residual intensity, then DMS is the only explanation.
  \item Therefore, we searched the ROSINA/DFMS data for observed signals of parent species that theoretically yield C$_2$H$_5$S on \textit{m/z}=61. Candidate molecules must contain at least two C atoms, five H atoms, and one S atom. S-bearing species with molecular masses larger than 61 Da observed on 15 March 2016 are S$_2$ and SO$_2$, both registered on \textit{m/z}=64, as well as CS$_2$ on \textit{m/z}=76, and their minor isotopologues. Those species are abundant and well-studied coma species \citep{leroy2015,calmonte2016}. However, they cannot produce the C$_2$H$_5$S fragment signal on \textit{m/z}=61 and therefore are not relevant. Up to \textit{m/z}=100, corresponding to the upper end of the mass range scanned on that day, no species which could contribute to the C$_2$H$_5$S fragment were observed. Apart from a few pure hydrocarbon species (C$_m$H$_n$), only krypton isotopes have been detected.
  \item The picture is different when the comet was close to its perihelion, beginning of August 2015. \citet{haenni2022} presented and discussed a data set collected on 3 August 2015, when the mass range was scanned up to \textit{m/z}=140 and the comet's coma was dusty. They reported the presence of a series of C$_m$H$_n$S-bearing species which shows that DFMS is indeed able to identify these species -- if they were present. The same authors derived upper limits in cases where distinction of the pure hydrocarbon and the S-bearing species with the same nominal mass was not possible as the exact masses are inseparable even at the mass resolution of DFMS. Intensities at the positions of the following C$_2$H$_5$S-bearing species have been observed: C$_3$H$_5$S, C$_3$H$_6$S, C$_3$H$_8$S, C$_4$H$_5$S, C$_4$H$_6$S, C$_6$H$_8$S, C$_3$H$_5$OS, C$_3$H$_6$OS, C$_3$H$_7$OS, C$_4$H$_8$OS, and C$_2$H$_6$S$_2$. We carefully checked if those species and their potential fragments, which are mostly \textit{m/z}=57, 64, 74, 76, 84, and 94 according to \citet{NIST}, are present also on 15 March 2016. However, the observed intensities are compatible with relevant and securely identified pure hydrocarbon species such as benzene. No sign of C$_2$H$_5$S-bearing species could be found.
  \item The exact mass scale and isotopic relative abundances also excludes contributions of C$_2$H$_6$S isotopologues. They are well below the uncertainty of the measurements. 
  %While the $^{34}$S- and the $^{13}$C-isotopologues, such as C$_2$H$_3$$^{34}$S and C$^{13}$CH$_4$S on \textit{m/z}=61, are clearly separable based on exact mass differences, the deuterated ones, such as C$_2$H$_3$DS on \textit{m/z}=61, potentially interfer because the exact mass difference is below the DFMS resolving power. However, via inference from the corresponding alkane ethane, where the D/H ratio has been determined as $(2.37\pm0.27)*10^{-3}$ by \citet{mueller2022}, we assume that the contribution of the fragments of C$_2$H$_5$DS can be neglected.
  %bei C2H5S ist das deuterierte ca. 1%. Das Fragment C2H5S ist ca, 37% von C2H6S, d.h. das deuterierte Fragment auf 62 dann 3.7 Promill, nicht relevant.
\end{enumerate}

Having ruled out contributions of heavier organosulfur species, we find DMS compatible (deviations between expected and observed intensities $<1\sigma$) and ethanethiol much less compatible (deviations between expected and observed intensities ~2-4$\sigma$) with ROSINA/DFMS data from 15 and 17 March 2016. Neither DMS nor possible precursors such as DMSP or other C-S-bearing molecules have been detected in the spacecraft background \citep{schlaeppi2010} and also reactions at the hot filament of the DFMS' ionization chamber are highly unlikely. Industrially, DMS is produced from the reaction of methanol and hydrogen disulfide over a hot aluminum oxide catalyst \citep{roy2000}. Both these molecules are common in comets, including comet 67P, where their abundance usually is on the order of 1$\%$ relative to water \citep{rubin2019b}. Nevertheless, the ionization chamber's open design, which does not allow pressure build-up, as well as the shielding of the hot filament impede chemical reactions. The ion source itself remained below 25$^{\circ}$C during the whole mission. We therefore argue that the observed DMS must be of cometary origin. Various lines of evidence, collected in \citet{haenni2022}, indicate a formation of volatile cometary organic molecules in the early Solar System history rather than on the comet itself.\\
\indent In order to estimate the abundance of DMS in comet 67P, we sum over all signals associated to DMS in the data set collected on 15 March 2016 between 03:30 and 04:14 UTC and give the result relative to methanol (CH$_3$OH). Methanol, a representative of complex organic molecules such as DMS itself, is not only common in comets but also in the interstellar medium (ISM), see, e.g., \citet{bockelee-morvan2000} or \citet{drozdovskaya2019}. From the analyzed spectra (Figures \ref{fig:spec} and \ref{fig:deconv}), we derive a relative abundance DMS/methanol = $(0.13\pm0.04)\%$ and methanethiol/methanol = $(0.9\pm0.27)\%$. The relatively large, estimated error accounts for the circumstance that that neither the ionization cross-sections nor the detector sensitivities of those molecules are known. Detailed information on the estimation of the error margins of abundances relative to methanol are in \citet{haenni2023}.\\
\indent Notably, abundances of specific molecules in comet 67P may expose pronounced time-variability. A time series for selected species has been published by \citet{laeuter2020}. This time-variability means that the relative abundance values reported for DMS in this work represent a snapshot. In Appendix \ref{sec:A1}, we delve into the time variability by comparing the data shown in Figures \ref{fig:spec} and \ref{fig:deconv} to other data sets collected around the time of 67P's outbound equinox in March 2016, in October 2014 \citep{calmonte2016}, while the spacecraft was orbiting the comet at a constant cometocentric distance of 10 km for roughly two weeks, and during its perihelion passage in early August 2015 \citep{haenni2022}. Due to contamination of the data by heavy organosulfur species, full deconvolution is only possible for the March 2016 data. Although the signals are varying slightly in intensity, as can be seen in Figure \ref{fig:A1}, the deduced C$_2$H$_5$S/C$_2$H$_6$S abundance ratios collected in Table \ref{tab:A1}, remain stable within error margins. Three out of four ratios are compatible with DMS as parent molecule within 1$\sigma$, while all four are off from the expectation value for ethanethiol by more than 3$\sigma$. Furthermore, the C$_2$H$_6$S/CH$_4$O abundance ratios were tabulated for the said data sets and demonstrate stability under similar conditions (March 2016 and October 2014). While demanding more detailed correlation studies similar to the work done by \citet{rubin2023}, this might hint at a shared origin of DMS and methanol not only in the comet but maybe even in the formational history. Additional pieces of the puzzle may be obtained from comparing the chemical complexity associated with cometary bulk ices \citep{rubin2019b,schuhmann2019a,schuhmann2019b} to that associated to dust \citep{altwegg2020,altwegg2022,haenni2022,haenni2023}.\\

\section{Conclusions}\label{sec:concl}
On Earth, DMS is a degradation product of DMSP, a sulfur-bearing compound produced by eukaryotic marine phytoplankton and a few higher plants \citep{andreae1983,pilcher2003}. It has been proposed \citep{seager2013} and employed \citep{madhusudhan2021,madhusudhan2023} as a unique biosignature in the search for biological activity on exoplanets. A thorough reanalysis of the high-resolution mass spectra collected by ROSINA/DFMS at comet 67P around the time of its outbound equinox, when \citet{calmonte2016} have previously reported a C$_2$H$_6$S signal, provides strong evidence for the presence of DMS in cometary matter. This is the first evidence of a non-anthropogenic abiotic formation pathway. However, whether or not the existence of cometary, abiotic DMS is relevant to exoplanetary atmospheres and hence has implications for the use of DMS as a biosignature, remains to be investigated in future work. \citet{fayolle2017} have hypothesized for the case of chloromethane (CH$_3$Cl), which is considered a biosignature attributed to oceanic bacteria on Earth \citep{seager2016} and has been detected in comet 67P, that exocomets impacting young exoplanetary surfaces may have delivered considerable amounts of material, including $6*10^5$ kg/yr of CH$_3$Cl under peak influx conditions \citep{brasser2016}. It is a topic of debate, though, if such a scenario is generally realistic or applicable to DMS and potentially other cometary species. Independent of DMS being employed as biosignature, we hope the detection of this molecule in pristine cometary matter encourages studies of abiotic formational pathways, for instance for dark molecular cloud conditions or protostellar environments. Given its respectable abundance relative to methanol derived from this work, DMS may be a promising candidate for ISM molecular searches (spectroscopic reference data are available from \citet{jabri2016} and \citet{ilyushin2020}). Laboratory work by \citet{mahjoub2017} with astrochemical ice analogs containing hydrogen sulfide and methanol showed a signal on \textit{m/z}=62 upon electron irradiation followed by temperature programmed desorption. However, the presented identification of DMS by these authors misses acknowledging the ambiguities created by unit mass resolution EI-MS and simultaneous sublimation of various sulfur-bearing molecules. Our results strongly motivate further investigation of the ISM sulfur-ice-chemistry as started by \citet{mahjoub2017}, instead focusing on the combination of small reduced sulfur and carbon carriers such as hydrogen sulfide and methane.\\

%%%%%%%%%%%%%%%%%%%%%%%%%%%%%%%%%%%%%%%%%%%%%%%%%%%%%%%%%%%%%%%%%%%%%%%

\acknowledgments

We gratefully acknowledge the work of the many engineers, technicians and scientists involved in the \textit{Rosetta} mission and in the planning, construction, and operation of the ROSINA instrument. Without their contributions, ROSINA would not have produced such outstanding data and scientific results. \textit{Rosetta} is an ESA mission with contributions from its member states and NASA. Work at the University of Bern was funded by the Canton of Bern and the Swiss National Science Foundation (200020\_207312). M.C. acknowledges support from NASA grants 80NSSC18K1280 and 80NSSC20K0651. J.D.K. acknowledges support from the Belgian Science Policy Office through PRODEX/ROSINA PEA 4000090020 and 4000107705. N.F.W.L. acknowledges support from the Swiss National Science Foundation (SNSF) Ambizione grant 193453. S.F.W. acknowledges support of the SNSF Eccellenza Professorial Fellowship PCEFP2\_181150. We are grateful to Daniel Kitzmann for enlightening discussions on the JWST observations of K2-18b.\\

%%%%%%%%%%%%%%%%%%%%%%%%%%%%%%%%%%%%%%%%%%%%%%%%%%%%%%%%%%%%%%%%%%%%%55

\appendix

\section{Appendix information}
\subsection{Time Variability of the Targeted Signals}\label{sec:A1}
\begin{figure}[ht]
   \centering
   \includegraphics[width=15cm]{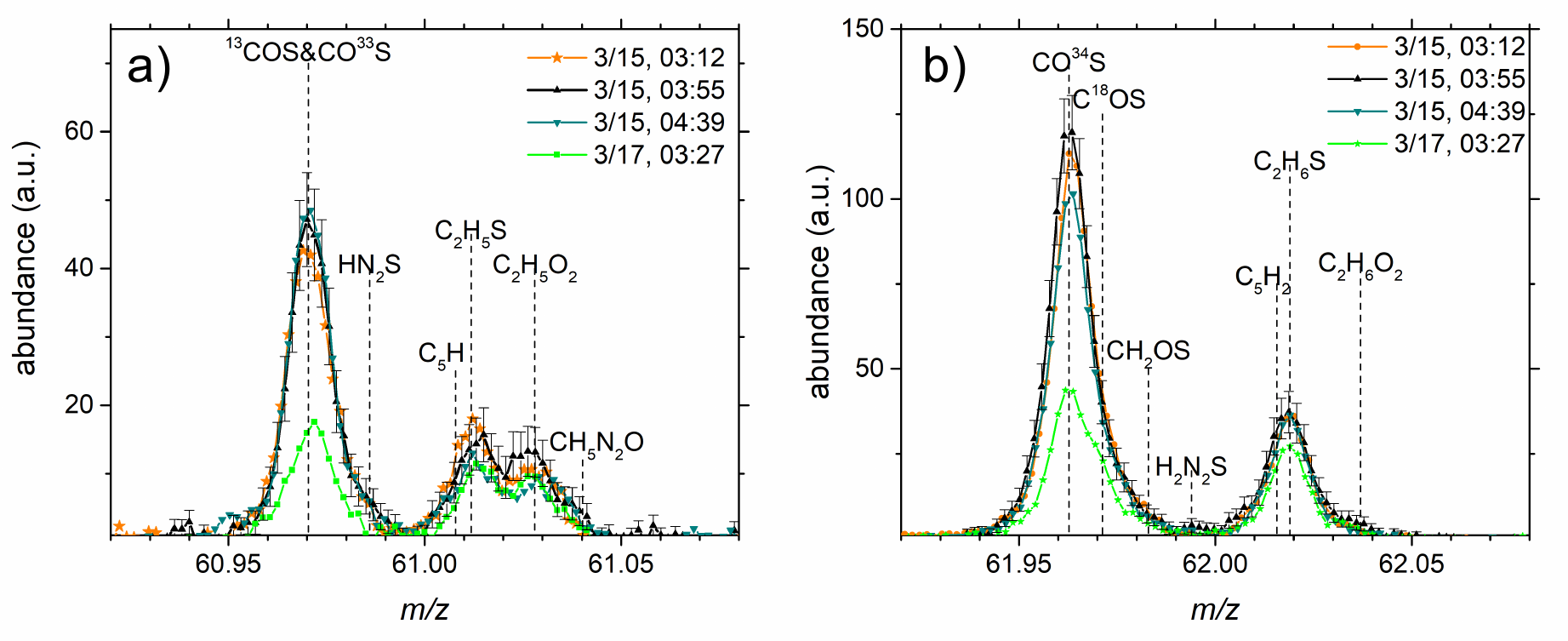}
      \caption{Time-variability of DFMS mass spectra collected around 67P's outbound equinox and for the set masses \textit{m/z}=61 (a) and \textit{m/z}=62 (b). The measured intensities per pixel are plotted with line points and in arbitrary units (a.u.). To improve the visibility, statistical uncertainties are indicated as error bars for the 3/15 03:55 UTC spectrum only.}
         \label{fig:A1}
\end{figure}

In order to investigate the time variability of the targeted S-bearing species, we have compared several \textit{m/z}=62 and \textit{m/z}=61 ROSINA/DFMS mass spectra mostly from mid-March 2016, see Figure \ref{fig:A1} and Table \ref{tab:A1}. In March 2016, comet 67P passed its outbound equinox, whith the spacecraft at a heliocentric distance of 2.58 au and at a cometocentric distance of approximately 13.5 km. Due to the rotation of the comet and the orbit of the spacecraft, the observable longitudes and latitudes were constantly changing on a short time scale, although DFMS has a large field of view of 20$^{\circ}$. In addition to the short-term variability, observed as a result of said changes in observational geometries or in cometary activity, there are other factors having an impact in the long-term. As the comet moved around the Sun and passed its dust and gas outflow activity maximum shortly after reaching perihelion in August 2015, the \textit{Rosetta} orbiter had to constantly readjust its relative position to the comet's surface for safety reasons. Sometimes, the distance also was changed for scientific reasons. For instance, shortly after the mid-March 2016 data was collected, the spacecraft retracted from the comet in favor of plasma studies to much greater cometocentric distances. As the local densities vary with $~1/r^2$, where $r$ is the cometocentric distance, relatively weak signals, $<1\%$ relative to water, are difficult to observe. For this reason, DMS was mainly observed when the spacecraft was close to the comet's surface at the beginning of the mission or around the analyzed equinox period. Its characteristic signal, i.e., C$_2$H$_6$S on \textit{m/z}=62, was just above detection limit when the comet was very active around perihelion, even though the spacecraft was at a relatively large cometocentric distance.\\
Table \ref{tab:A1} summarizes the spacecraft's relative position together with the C$_2$H$_5$S/C$_2$H$_6$S and the C$_2$H$_6$S/CH$_4$O abundance ratios for the four spectra shown in Figure \ref{fig:A1} plus two additional ones. For the latter two, from 10 October 2014 \citep{calmonte2017} and 3 August 2015 \citep{haenni2022}, no full analysis can be done as, in both cases, heavier C$_n$H$_m$S species and potentially even C$_5$H are contaminating the C$_2$H$_5$S signal. The derived C$_2$H$_5$S/C$_2$H$_6$S ratios thus have a large uncertainties. From the listed abundance ratios we can conclude two things: First, three out of four C$_2$H$_5$S/C$_2$H$_6$S abundance ratios are compatible with DMS within 1$\sigma$, while for ethanethiol all are off by more than 3$\sigma$. Note that for DMS, the expected C$_2$H$_5$S/C$_2$H$_6$S ratio is 33\%, while for ethanethiol it is 15\% \citep{NIST}. Second, the C$_2$H$_6$S/CH$_4$O abundance ratios, where CH$_4$O corresponds to the methanol molecular ion, are surprisingly constant too, except for the one in August 2015 when there was a lot of sublimation also from dust and the cometary activity was highly variable. In October 2014 and March 2016, however, the comet was at a comparable heliocentric distance of around 3 au and the activity was more stable and rather comparable. The respective C$_2$H$_6$S/CH$_4$O abundance ratios are identical within 1$\sigma$ error margins. This is an indication for DMS and methanol to sublimate from the same layer in the cometary nucleus, and possibly even hints at a common origin. However, more detailed correlation studies are needed to support this hypothesis.

\begin{table}[h!]
    \centering
    \begin{tabular}{ccccc}
        \hline
        Time (UTC) & Longitude ($^{\circ}$) & Latitude ($^{\circ}$) & C$_2$H$_5$S/C$_2$H$_6$S (\%) & C$_2$H$_6$S/CH$_4$O (\%) \\
        \hline
        15 March 2016, 03:12 &   29.08 & -14.63 & 43 $\pm$ 6 & 0.13 $\pm$ 0.04 \\
        15 March 2016, 03:55 &    7.37 & -17.10 & 31 $\pm$ 6 & 0.13 $\pm$ 0.04 \\
        15 March 2016, 04:40 &  -14.40 & -19.59 & 36 $\pm$ 6 & 0.14 $\pm$ 0.04 \\
        17 March 2016, 03:27 & -140.18 &  21.42 & 40 $\pm$ 7 & 0.11 $\pm$ 0.04 \\
        \hline
        10 October 2014, 17:17 & -13.45 &  6.59 & 100 $\pm$ 20 & 0.12 $\pm$ 0.04 \\
        3 August 2015, 16:48 & -31.09 &  -13.78 & 95 $\pm$ 30 & 0.26 $\pm$ 0.06 \\
        \hline
    \end{tabular}
    \caption{Spacecraft relative position (longitude and latitude) and C$_2$H$_5$S/C$_2$H$_6$S and C$_2$H$_6$S/CH$_4$O abundance ratios for the four spectra shown Figure \ref{fig:A1} plus two additional spectra from 10 October 2014 \citep{calmonte2016} and 3 August 2015 \citep{haenni2022}, respectively.}
    \label{tab:A1}
\end{table}

\bibliography{literature}{}
\bibliographystyle{aasjournal}



\end{document}